\documentclass[reprint, aps, prl, amsmath, amssymb, superscriptaddress]{revtex4-2}
\usepackage{amsmath}

\setcitestyle{super,open={},close={}}

\usepackage{multirow}
\usepackage{siunitx}
\usepackage{dsfont}

\usepackage{ragged2e}
\usepackage{caption, sansmath}

\DeclareCaptionFont{sansmath}{\sansmath}
\DeclareCaptionFormat{custom}
{%
    \textbf{#1#2}{\justifying #3}
}
\DeclareCaptionLabelFormat{lc}{Fig.~#2}
\usepackage{upgreek}
\usepackage{lmodern} 
\usepackage[bookmarks, colorlinks=false]{hyperref}
\usepackage{mathtools}
\usepackage{bbold}
\usepackage{bm}
\hypersetup{colorlinks=true, allcolors=blue}

\begin{document}

\renewcommand{\vec}[1]{\boldsymbol{#1}}

\renewcommand{\tensor}[1]{\vec{\mathrm{#1}}}

\hyphenation{milli-dpa}

\title{Fast low-temperature irradiation creep driven by athermal defect dynamics}

\author{Alexander Feichtmayer}
	\affiliation{Max Planck Institute for Plasma Physics, Boltzmannstr.\,2, 85748 Garching, Germany}
    \affiliation{Technical University Munich, Boltzmannstr. 15, 85748 Garching, Germany}
 
\author{Max Boleininger}
	\email{max.boleininger@ukaea.uk}
	\affiliation{UK Atomic Energy Authority, Culham Centre for Fusion Energy, Oxfordshire OX14 3DB, United Kingdom}

\author{Johann Riesch}
	\affiliation{Max Planck Institute for Plasma Physics, Boltzmannstr.\,2, 85748 Garching, Germany}
 
\author{Daniel R. Mason}
	\affiliation{UK Atomic Energy Authority, Culham Centre for Fusion Energy, Oxfordshire OX14 3DB, United Kingdom}

\author{Luca Reali}
	\affiliation{UK Atomic Energy Authority, Culham Centre for Fusion Energy, Oxfordshire OX14 3DB, United Kingdom}

\author{Till Höschen}
	\affiliation{Max Planck Institute for Plasma Physics, Boltzmannstr.\,2, 85748 Garching, Germany}

\author{Maximilian Fuhr}
	\affiliation{Max Planck Institute for Plasma Physics, Boltzmannstr.\,2, 85748 Garching, Germany}
    \affiliation{Technical University Munich, Boltzmannstr. 15, 85748 Garching, Germany}

 \author{Thomas Schwarz-Selinger}
	\affiliation{Max Planck Institute for Plasma Physics, Boltzmannstr.\,2, 85748 Garching, Germany}

 \author{Rudolf Neu}
	\affiliation{Max Planck Institute for Plasma Physics, Boltzmannstr.\,2, 85748 Garching, Germany}
    \affiliation{Technical University Munich, Boltzmannstr. 15, 85748 Garching, Germany}

\author{Sergei L. Dudarev}
	\affiliation{UK Atomic Energy Authority, Culham Centre for Fusion Energy, Oxfordshire OX14 3DB, United Kingdom}

\date{\today}
\begin{abstract}
\noindent The occurrence of high stress concentrations in reactor components is a still intractable phenomenon encountered in fusion reactor design. We observe and quantitatively model a non-linear high-dose radiation mediated microstructure evolution effect that facilitates fast stress relaxation in the most challenging low-temperature limit. \textit{In situ} observations of a tensioned tungsten wire exposed to a high-energy ion beam show that internal stress of up to {2\,GPa} relaxes within minutes, with the extent and time-scale of relaxation accurately predicted by a parameter-free multiscale model informed by atomistic simulations. As opposed to conventional notions of radiation creep, the effect arises from the self-organisation of nanoscale crystal defects, athermally coalescing into extended polarized dislocation networks that compensate and alleviate the external stress.
\end{abstract}
\maketitle

\noindent 
The central engineering challenge in realising commercially viable fusion power-generating reactors lies in the degradation of materials in components surrounding the fusion plasma\cite{knaster2016materials}. Structural components inside the vacuum vessel are exposed to extreme conditions\cite{coenen2015materials}, involving high magnetic fields, gravitational loads, plasma disruptions, and unprecedented levels of neutron and gamma radiation\cite{reali2023intense} that continuously generate microscopic defects, leading to severe degradation of properties of materials critical to their intended function. Furthermore, spatio-temporal variations in irradiation exposure and temperature induce stress concentrations of sufficient magnitude to threaten the integrity of load-bearing or physical barrier\cite{wu2016identification,Reali2022, Griffiths2023} components.

The tendency of metals to undergo steady viscoplastic deformation when subjected to mechanical stress at elevated temperatures, known as {\it creep}, is of pivotal significance to the structural integrity of a reactor. Irradiation by high-energy particles significantly accelerates creep, causing deformation rates orders of magnitude higher than thermal creep under the otherwise equivalent stress\cite{gilbert1977dependence, was2017irradiation}. However, mechanistic understanding of the phenomenon proves elusive, partly because it is challenging to design experiments enabling accurate measurements under simultaneous high irradiation fluence, mechanical, and thermal loads.

In particular, the role of temperature remains undetermined. Conventional argument suggests that creep rates increase with temperature due to the enhanced thermal diffusion of microstructural defects, mediating the viscoplastic flow. In the apparent contradiction with this argument, measurements of irradiation creep in steels\cite{grossbeck1991low} show that at low temperatures the irradiation creep occurs at an anomalously {\it high} rate, decreasing as a function of temperature, only to increase again at higher temperatures. Low temperatures, and large temperature and radiation exposure gradients are expected in the actively cooled reactor components, making a rational model for the phenomenon a requirement for an expert fusion reactor design.

\begin{figure*}[t]
\includegraphics[width=.95\textwidth]{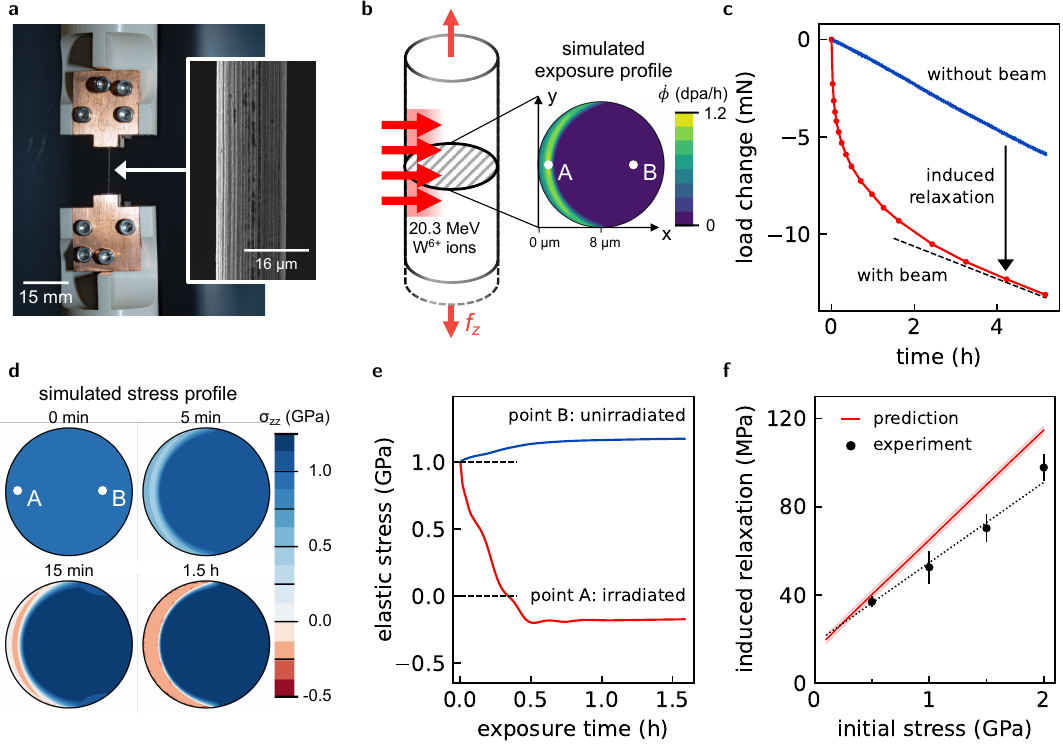}
\caption{\textbf{Time-resolved measurement of stress relaxation under irradiation.} \textbf{a,} A thin tungsten wire of length {15\,mm} and diameter {16\,\textmu m} is loaded with tensile stress, after which a central wire section of {4\,mm} length is exposed to a beam of {20.3\,MeV} W$^{6+}$ ions whilst monitoring the force required to maintain the initial elongation. \textbf{b,} The ion beam damages a thin surface layer of about {2\,\textmu m} thickness, as seen in the simulated exposure profile in the cross-section. \textbf{c,} Upon exposure of the wire to the beam, a rapid relaxation of load is measured. \textbf{d-e,} While the wire is initially under uniform tensile stress, a multi-scale simulation of the developing internal stress reveals that in the irradiated surface layer (A), the tensile stress relaxes completely and even becomes compressive, while in the unexposed region deeper in the wire (B), the tensile stress increases to balance the expansion of the irradiated layer. \textbf{f,} The final magnitude of induced stress relaxation is proportional to the magnitude of the initial tensile stress, suggesting that the relaxation is strongly biased by the external stress. Error bars indicate the standard error over multiple experiments. The shaded region indicates the 3$\sigma$-confidence interval of the simulation prediction, with uncertainty originating from the stochasticity of the atomistic data underlying the surrogate model.
\label{fig:1}
}
\end{figure*}

In this study, we develop a comprehensive quantitative treatise of the anomalous low-temperature irradiation creep phenomenon by combining custom-designed experiments with \textit{in situ} measurement capabilities and a parameter-free multiscale model. Fig.~\ref{fig:1} illustrates a time-resolved observation of stress relaxation in a {16\,\textmu m} thin tungsten wire  -- thinner than a human hair -- under exposure to highly energetic ions. Initially, the wire is stretched until a specified tensile load is reached. Subsequently, the wire is exposed to the ion beam while the force required to maintain its length is monitored. Despite the wire remaining at a temperature where no thermally-driven creep occurs, our observations reveal an initial rapid relaxation of the force occurring within minutes, followed by slower relaxation plateauing over a few hours. Significantly, in Fig.~\ref{fig:1}f we observe the extent of relaxation to be linearly proportional to the external load, which is in remarkable qualitative and quantitative agreement with predictions derived from an \textit{in silico} replication of the experiment through a parameter-free model. Our analysis shows that the relaxation is non-linear and involves the coarsening of microscopic defects under external stress. Since only {4\,\%} of the wire materials is exposed to high-energy ions, the extent of the measured relaxation suggests that locally in the irradiated surface layer, the applied tensile loads of up to {2\,GPa} relax entirely and solely through the anomalous fast low-temperature irradiation creep.

\

\noindent\textbf{\sffamily Time-resolved stress relaxation measurement}\\
To measure the time-resolved irradiation-induced stress relaxation, it is necessary to select an appropriate high-energy particle source. Neutrons penetrate deep into materials, allowing for testing of bulk samples, but the conditions in a typical materials test reactor rarely permit {\it in situ} mechanical testing. Neutron activation of materials also complicates handling, with low dose rate implying long exposure times. For these reasons, we chose irradiation by high-energy heavy ions. These are readily available, can be produced in the form of intense beams, and cause no activation. It is also possible to incorporate a mechanical testing setup into an ion accelerator beam line.

As high-energy heavy ions have a penetration depth of only a few \SI{}{\um}, the exposed zone should span many grains in order to replicate bulk-like irradiation conditions. We selected potassium-doped ({60–75\,ppm}) cold-drawn tungsten wires, as tungsten is the prime candidate material for plasma-facing components due to its high melting point and high thermal conductivity. The wires are industrially produced with diameters as small as \SI{16}{\um} and feature nanoscale grains. No thermal creep is expected, as only low homologous temperatures are reached ($T<0.1\, T_\mathrm{melt}$). Extensive studies of mechanical properties\cite{Riesch_2017, fuhr2023rate}, deuterium retention\cite{Kaercher2021}, and post-irradiation properties\cite{Riesch_2022} of this material are available. To prevent the implantation of impurities, we employ a beam of \SI{20.3}{MeV} W$^{6+}$ ions. These ions penetrate about \SI{2}{\um} into tungsten, exposing about \SI{16}{\percent} of the wire cross-section.

\begin{figure}[t]
\includegraphics[width=\columnwidth]{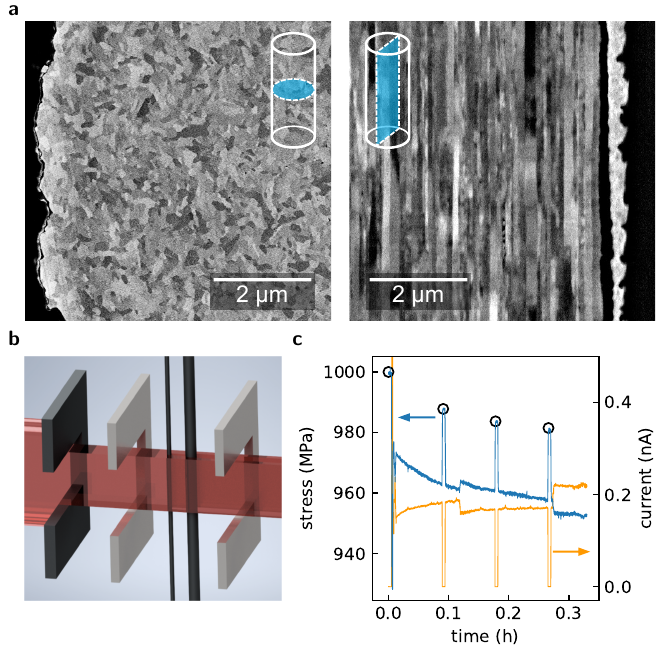}
\caption{\textbf{a,} Microstructure of the {16\,\textmu m} tungsten wire. Left: cross-section images taken using a scanning electron microscope (SEM) in backscattered electron contrast mode. Right: longitudinal section using secondary electron contrast. \textbf{b,} Schematic overview of the ion beam path during the experiment. From left to right: beam limiting aperture, first suppression electrode, {16\,\textmu m} wire sample, {102\,\textmu m} measuring wire, second suppression electrode.  \textbf{c,} Tensile stress (blue) and ion current (orange) during the experiment. Black circles indicate the points where experimental measurements were performed between the irradiation intervals.
\label{fig:2}
}
\end{figure}

To enable a time-resolved measurement of the stress relaxation phenomenon, a dedicated experiment comprising a tensile testing machine mounted in a vacuum vessel was set-up at a beamline of the \SI{3}{MV} tandem accelerator at the Max Planck Institute for Plasma Physics. The \SI{15}{mm} long wire samples are mounted between two crossheads of the machine and mechanically tensioned to a uniaxial stress ranging from \SI{0.5}{GPa} to \SI{2.0}{GPa}, which is still in the elastic region. While the strain is kept constant, a central wire section of \SI{4}{mm} length is exposed to the ion beam. Since almost all of the impacting ions' kinetic energy is absorbed in the wire, the sample's temperature rises to at most \SI{100}{\celsius}. To correct for the influence of the resulting thermal expansion on the force measurement, the beam is periodically turned off at intervals of \SIrange{5}{30}{min}. The sample cools down over these intervals, allowing for the measurement of the actual force drop. The ion fluence is determined by measuring the electric current flowing through the wire, from which the dose in ``displacements per atom'' (dpa)\cite{norgett1975proposed} is computed using the number of displacements per colliding ion from the SRIM software\cite{ziegler2004srim}. The samples are exposed to a peak dose of up to \SI{6}{dpa} over a typical \SI{6}{h} duration experiment. The measured forces are adjusted for shifting and relaxation of the load cell and measurement setup, and further steps are taken to improve the signal-to-noise ratio of the measured ion current,  see the Methods sections. Figure \ref{fig:2}b schematically illustrates the irradiation part of the experiment. 


\

\noindent\textbf{\sffamily Insights from in silico testing}

\noindent While the extent of relaxation in the irradiated volume can be numerically estimated, such estimates do not account for the transient and heterogeneous nature of the stress field developing over the course of irradiation. Similarly, even if microstructural evolution were precisely characterised using imaging techniques, without understanding the physical principles governing it, this insight would not be transferable to complex geometries involving spatially or temporally varying stress fields.

\begin{figure}[t]
\includegraphics[width=.93\columnwidth]{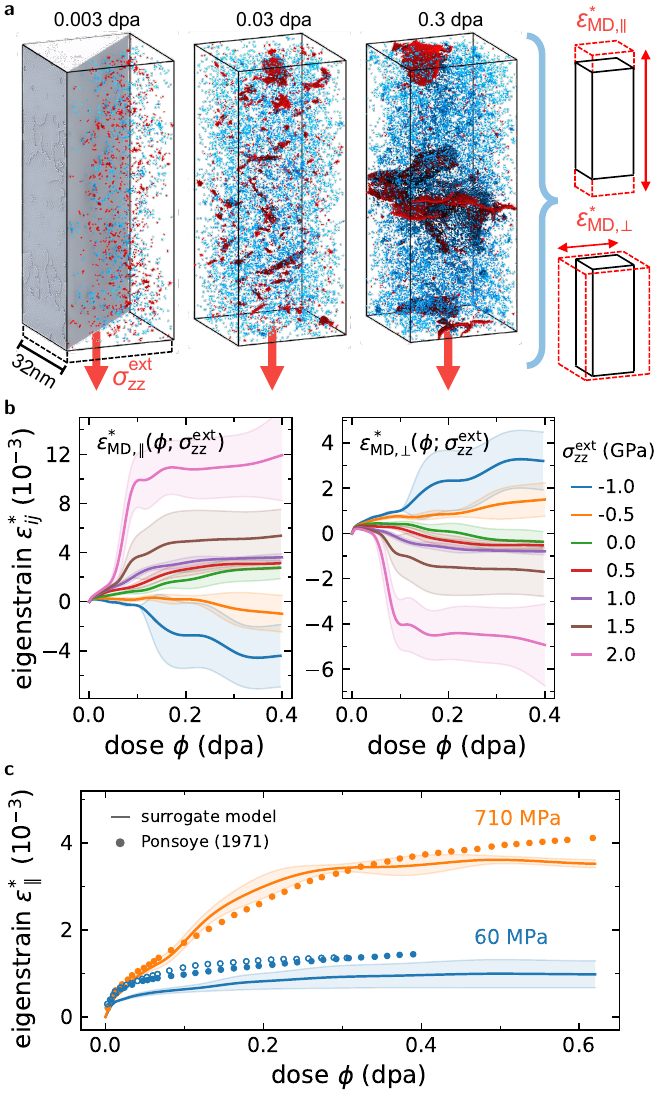}
\caption{\textbf{Data-driven surrogate model.}
\textbf{a,} Atomistic simulations generate data on deformation eigenstrains for a representative volume element under irradiation and external uniaxial stress. Radiation damage is simulated by assigning randomly chosen atoms in a tungsten crystal (white cutaway) recoil energies, drawn from scattering events with {20.3\,MeV W$^{6+}$} ions, leading to initiation of collision cascades that result in the formation of nanoscale interstitial (red) and vacancy (blue) defects, shown here for {1\,GPa}. Eigenstrains are extracted in directions parallel ($\parallel$) and perpendicular ($\perp$) to the uniaxial stress direction.
\textbf{b,} A surrogate model is matched to the data using maximum likelihood estimation. Solid lines and shaded areas indicate the mean value and standard deviation, respectively, of eigenstrains from five repeated simulations per stress value. 
\textbf{c,} The surrogate model, when adjusted to isotropic expansion at zero stress (see text), is in agreement with \textit{in situ} measurements of elongation strain of tungsten irradiated by fission fragments at {20\,K} under tension.
\label{fig:3}
}
\end{figure}

We have opted to develop a virtual representation of the wire experiment based on a small number of precisely defined and tested theoretical principles. The stress relaxation mechanism is described by a surrogate model, derived from the data obtained directly from atomistic simulations, with no additional parameters adjusted to fit the experimental result. In this way, the model has predictive capabilities over a well-defined range of experimental conditions. By embedding the surrogate model into a finite element representation of the wire, we bridge the gap between atomic and micro-mechanical scales, arriving at a ``digital shadow'' of the experiment. Atomistic simulations also provide full and precise information about the microstructure, enabling the {\it in silico} identification of the mechanisms responsible for the observed relaxation.

Molecular dynamics (MD) simulations have been used to successfully predict  properties of materials exposed to irradiation without the need for adjustable parameters\cite{mason2019relaxation,  hirst2022revealing, boleininger2023microstructure}. While MD simulations enable direct and parameter-free analysis of response of radiation effects to thermal and mechanical loads, it would require over \SI{e12}{atoms} just to represent a \SI{1}{\micro\meter} long wire section. The largest simulation reported in literature involved \SI{2e9}{atoms}\cite{zepeda2021atomistic}. Consequently, with MD we cannot explicitly simulate the heterogeneous dose profile attenuating on the micrometer scale. In what follows, we employ MD to describe the behaviour of a representative volume element (RVE) under irradiation. The simulation operates as a black box; given an external homogeneous stress, we obtain data on dimensional changes of an RVE during irradiation.

In our atomistic simulations, see Fig.~\ref{fig:3}, the initially pristine tungsten single crystals were exposed to atomic recoils representative of those initiated by \SI{20.3}{MeV} W$^{6+}$ ions, while maintaining a specified external uniaxial load $\sigma_\mathrm{zz}^\mathrm{ext}$. These recoils lead to the generation of self-similar cascades of atomic displacements, ultimately resulting in the formation of crystal defects that cause dimensional changes of the RVE, here described as eigenstrains parallel and perpendicular to the uniaxial stress direction, $\varepsilon^*_{\parallel}$ and $\varepsilon^*_{\perp}$, respectively. While the simulated dose-rate of {20\,dpa/\textmu s} is nine orders of magnitude higher than in experiment, at a temperature of at most {$100^\circ$C}, vacancies remain immobile with a mean migration time of {1\,year}. Interstitials, on the other hand, are mobile, driven by thermal {\it and} stress fluctuations. As a result, microstructural evolution follows an effectively athermal pattern, both in experiment and simulations\cite{boleininger2023microstructure}.

Because of the discrete nature of atomic recoils, MD simulations inherently generate discrete datasets of eigenstrain $\{(\phi_0, \tensor{\varepsilon}^*_0), (\phi_1, \tensor{\varepsilon}^*_1),\, \cdots, (\phi_N, \tensor{\varepsilon}^*_N)\}$ over dose $\phi_i$ in dpa. Moreover, independent MD simulations for the same uniaxial stress yield different eigenstrain datasets due to the stochasticity inherent to collision cascades\cite{Sand2013} and dislocation motion\cite{rizzardi2022mild, proville2024unravelling}. Since these eigenstrains are intended for the use in a homogenised finite element model, we require a surrogate model representing the expectation value $\mu$ and variance $s^2$ of eigenstrains as continuous functions of dose and stress. We express the probability to draw a parallel or perpendicular eigenstrain component $(\phi_i$, $\varepsilon^*_i)$ from a normal distribution $\mathcal{N}(\mu, s^2)$ by
\begin{equation}
    p(\vec{\theta} | \varepsilon^*_i) = \frac{1}{\sqrt{2\pi}s(\phi_i; \vec{\theta})} \exp\left(
        -\frac{\left(\varepsilon^*_i - \mu(\phi_i; \vec{\theta})\right)^2}{2 s^2(\phi_i; \vec{\theta})}
    \right),
\end{equation}
where $\mu(\phi; \vec{\theta})$ and $s(\phi; \vec{\theta})$ are cubic spline functions with unknown knot heights $\vec{\theta}$. The likelihood to draw the entire eigenstrain data for a common external stress $\sigma_\mathrm{zz}^\mathrm{ext}$ from the model is
\begin{equation}
    \mathcal{L}(\vec{\theta} | \vec{\varepsilon}^*) = \prod_{i=1}^N p(\vec{\theta} | \varepsilon^*_i).
\end{equation}
By maximising the log-likelihood over $\vec{\theta}$,
\begin{equation}
    \hat{\vec{\theta}} = \underset{\vec{\theta} \in \Theta}{\arg\max} \log \mathcal{L}(\vec{\theta} | \vec{\varepsilon}^*),
\end{equation}
we obtain a maximum likelihood estimate for $\hat{\vec{\theta}}$ best matching the data. This process is applied to $\varepsilon_\parallel$ and $\varepsilon_\perp$ components, independently for each simulated stress.  Finally, $\mu(\phi; \vec{\theta})$ is interpolated in stress-space using cubic splines, resulting in continuous representations of the eigenstrains $\varepsilon^*_{\mathrm{MD},\parallel}(\phi,\sigma_\mathrm{zz}^\mathrm{ext})$ and $\varepsilon^*_{\mathrm{MD},\perp}(\phi,\sigma_\mathrm{zz}^\mathrm{ext})$. From these, the diagonal eigenstrain tensor is formed with $\tensor{\varepsilon}^*_\mathrm{MD}(\phi, \sigma_\mathrm{zz}^\mathrm{ext}) = \mathrm{diag}(\varepsilon^*_{\mathrm{MD},\perp}, \varepsilon^*_{\mathrm{MD},\perp}, \varepsilon^*_{\mathrm{MD},\parallel})$. The eigenstrain input data are obtained from five repeated MD simulations for each stress $\sigma_\mathrm{zz}^\mathrm{ext}$ between \SI{-1.0}{GPa} and \SI{2.0}{GPa} in increments of \SI{0.5}{GPa}. Additional details are given in the Methods section. 

It remains to formulate a model for the incremental accumulation of eigenstrain $\tensor{\varepsilon}^*$ during irradiation. In experiment, the wire begins in a uniform stress state that becomes non-uniform upon exposure to the spatially attenuating ion beam. The surrogate model, however, describes dimensional changes of a RVE irradiated under a homogeneous external stress. The eigenstrain induced by a complex microstructure generally comprises contributions with varying degrees of reversibility in response to a changing stress, and the same is expected to hold true for microstructures formed by irradiation. Atomistic simulations show that eigenstrain is largely irreversible; a momentary change in external stress has little effect on the eigenstrain. A similar observation was recently reported for aluminium irradiated under stress\cite{da2023evidence}. Fig.~\ref{fig:3}a suggests that at high dose, interstitial defects preferentially arrange themselves into extended structures, such as dislocation networks or new crystal planes\cite{boleininger2022volume}, which cause dimensional change that is no longer affected by external stress. Motivated by this observation, we assume that eigenstrains accumulate irreversibly over the course of irradiation, with the accumulation driven by the surrogate model evaluated for the instantaneous and local elastic stress $\tensor{\sigma}^\mathrm{el}(\vec{r},t)$ and dose $\phi(\vec{r},t)$ fields:
\begin{equation}
\frac{\mathrm{d}}{\mathrm{d}t}\tensor{\varepsilon}^*(\vec{r},t) 
= \frac{\mathrm{d}}{\mathrm{d}t}
    \tensor{\varepsilon}_\mathrm{MD}^*[\phi, \tensor{\sigma}^\mathrm{el}](\vec{r},t),
\end{equation}
where $\tensor{\sigma}^\mathrm{el} = \tensor{\sigma} - \tensor{\sigma}^*$ is the elastic stress, with $\tensor{\sigma}^* \equiv \tensor{C} : \tensor{\varepsilon}^*$ using the elastically isotropic stiffness tensor $\tensor{C}$ of tungsten\cite{HirthLothe}.

Lastly, because elastic fields have long range, variations in the local stress are implicitly dependent on the spatially heterogeneous stress field involving the entire wire. The gap in length scales between the nanoscale RVE and the microscale wire is bridged by representing the eigenstrain as a body-force acting in a linear elasticity framework\cite{Reali2022}, expressed as $\vec{f}^* = -\vec{\nabla}\cdot \tensor{\sigma}^*$:
\begin{equation}\label{eq:elastostatic}
\vec{\nabla}\cdot \tensor{\sigma}(\vec{r},t) = \vec{\nabla}\cdot \tensor{\sigma}^*(\vec{r},t).
\end{equation}
We solve for the stress tensor field $\tensor{\sigma}$ using the finite element method, with the time propagation scheme described in the Methods section. 

\begin{figure}[t]
\includegraphics[width=\columnwidth]{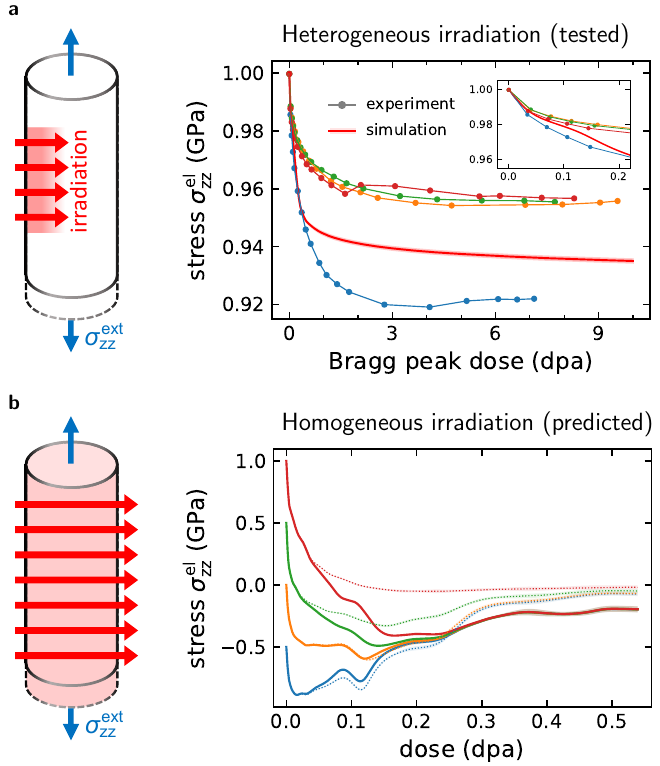}
\caption{\textbf{Stress relaxation under heterogeneous and homogeneous irradiation.} \textbf{a,} Comparison of four irradiation-induced stress relaxation experiments to the parameter-free simulation using the virtual wire model under an initial uniaxial tensile stress of {1\,GPa}. The model predicts a small degree of relaxation, in agreement with experiment, due to the small volume fraction of the irradiated volume. \textbf{b,} Predicted stress relaxation curves for a hypothetical homogeneous irradiation scenario under various initial uniaxial tensile and compressive stresses. Most of the relaxation is predicted to occur within the first {0.3\,dpa}.
\label{fig:4}
}
\end{figure}

In Fig.~\ref{fig:4}a, we compare \textit{in situ} measurements and \textit{in silico} predictions of stress relaxations for the initial applied stress of {1\,GPa}. Over time, we observe in the stress profile shown in Fig.~\ref{fig:1}d, the emergence of mutually counterbalancing compressive and tensile elastic stresses in the irradiated and unirradiated regions, respectively. The material exposed to irradiation deforms plastically to accommodate the external tensile stress, resulting in local stress relaxation. This deformation likewise gives rise to a long-ranged elastic tensile stress field extending well into the unexposed region, increasing the tensile stress beyond its initial value, as depicted in Fig.~\ref{fig:1}e.

The measured and predicted final relaxations for the various initial stresses are shown in Fig.~\ref{fig:1}f. The predictions are in good qualitative agreement with experimental results, with an overestimation of approximately {20\,\%}. The amount of relaxation appears to obey a linear relationship with the initial external stress $\sigma_\mathrm{zz}^\mathrm{ext}$. Specifically, the simulation data follows the curve given by $\Delta\sigma^\mathrm{sim} = 14.9\,\mathrm{MPa} + 0.050\, \sigma_\mathrm{zz}^\mathrm{ext}$, whereas the experimental data follows $\Delta\sigma^\mathrm{exp} = 18.1\,\mathrm{MPa} + 0.036\, \sigma_\mathrm{zz}^\mathrm{ext}$. Assuming that at high dose, the elastic stress relaxes completely within the irradiated region, the fraction of relaxed stress should be close to the irradiated volume fraction $V_\mathrm{irr}/V = 0.04$; this is in agreement to the proportionality constants of the linear relations. Bearing this in mind, the major source of uncertainty in our virtual wire model originates not from the stochastic uncertainty of atomistic data, but rather from the dose profile determining the irradiated volume fraction.

We observe that the irradiated region relaxes to a lightly compressive state of {$-200$\,MPa} instead of an entirely stress-free state. This is a finite-size effect originating from atomistic simulations, where we used a simulation cell with a shape elongated in the $z$-direction (32$\times$32$\times$80\,{nm$^3$}). In a stress-free simulation, at low dose, the interstitial loops are in prismatic orientation, with their habit planes oriented perpendicular to the Burgers vectors, thus giving rise to isotropic volumetric swelling. As the loops grow under continued irradiation, they interact with their own periodic image across the shorter simulation cell sides, eventually forming extended dislocation networks or new complete crystal planes; as a result the initially isotropic expansion becomes preferentially polarised in $z$-direction. For the same reason, the final induced relaxation, shown in Fig.~\ref{fig:1}f, does not vanish at zero initial stress, as tensile eigenstrain develops under irradiation even in the absence of external stress. The same effect is also observed in experiment, where the grains are similarly elongated along $z$ ($\sim$50$\times$50$\times$1000\,{nm$^3$}), suggesting a genuine irradiation growth phenomenon caused by the anisotropic grain microstructure. 

Finally, to assess the potential beneficial effect of fast low-temperature stress relaxation under a realistic reactor operating scenario, we applied the model to a hypothetical case involving a homogeneously irradiated wire, illustrated in Fig.~\ref{fig:4}b. This scenario is experimentally inaccessible with heavy ion irradiation due to their limited implantation range and is, instead, more representative of neutron irradiation. Using the virtual wire, we can also predict relaxation under a compressive stress, a condition that cannot be tested in our experiments where the wire would buckle. Our predictions exhibit the rapid complete relaxation of the initial stress within {0.3\,dpa} of exposure. 
In the same figure, we also show relaxation as predicted by a surrogate model constrained to expand isotropically under stress-free conditions (dotted lines), which is more representative of tungsten with isotropic grain microstructure.

\begin{figure*}[t]
\includegraphics[width=.95\textwidth]{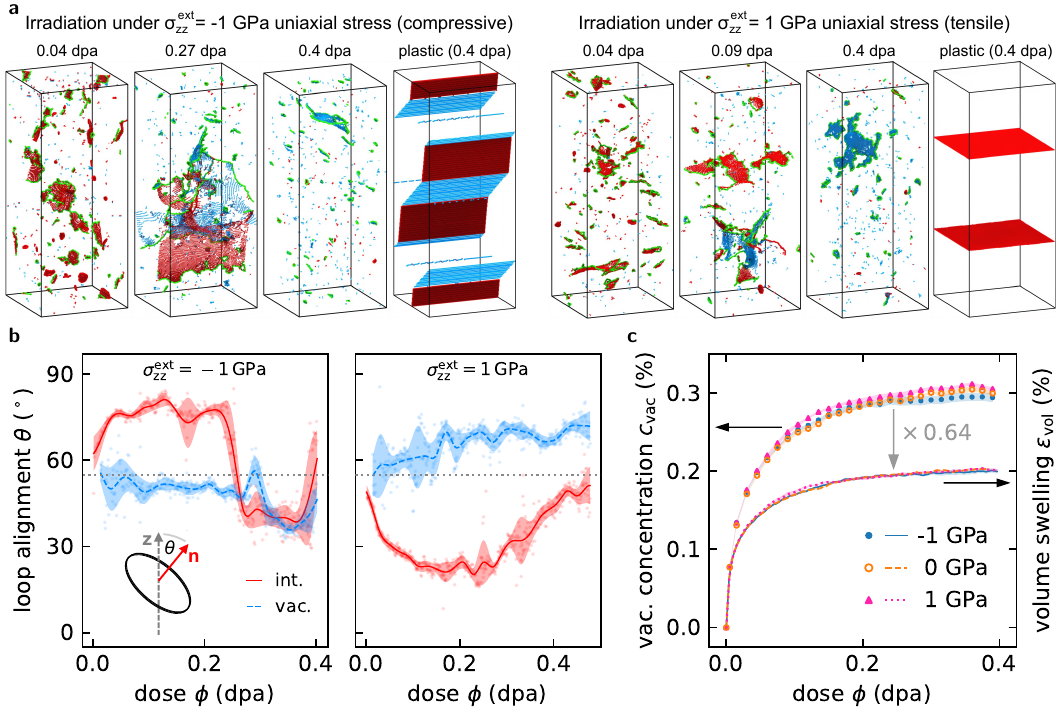}
\caption{\textbf{Anomalous low-temperature irradiation creep.} \textbf{a,} Irradiation creep is driven by the coarsening and coalescence of highly mobile interstitial clusters. At low dose, irradiated microstructure contains mostly interstitial-type (red) and rarely vacancy-type (blue) dislocation loops, as well as dispersed vacancies. With further exposure, the interstitial-type loops grow and coalesce, eventually forming a complex, system-spanning dislocation network of mixed character which enables plastic deformation of the crystal. Shown are defect clusters (N$>$2) and dislocations (1/2$\left\langle 111 \right\rangle$: green, $\left\langle 100 \right\rangle$: red) as identified with the Wigner-Seitz and DXA\cite{stukowski2009visualization} methods, respectively. The reference crystal for Wigner-Seitz analysis is chosen to minimise the number of point defects, allowing for direct visualisation of plastic deformation. \textbf{b,} Interstitial and vacancy-type loops orient themselves into opposing directions relative to the external uniaxial stress direction, which leads to a stress-bias in the eventual formation of the dislocation network. Shown are mean alignments for a given dose (dots) and spline-interpolated values (line) with standard error (shaded). The dashed line indicates prismatic alignment. \textbf{c,} The extent of volume swelling is independent of external stress and can be well estimated by 0.64 times the vacancy concentration, with the numerical factor arising from comparison of interstitial and vacancy defect relaxation volumes, see text. Shaded areas indicate the standard error over five repeated simulations.
\label{fig:5} 
}
\end{figure*}

\

\noindent\textbf{\sffamily Discussion}

\noindent Inspecting the atomistic simulations underpinning the surrogate model, we have identified three distinct mechanisms contributing to stress relaxation occurring over some distinct dose and stress intervals. Up to the dose of about {0.05\,dpa}, dislocations form a dispersed microstructure, primarily consisting of $1/2\left\langle 111 \right\rangle$-type interstitial loops shown in Fig.~\ref{fig:5}a. In the absence of external stress, the prismatic loops generate isotropic volumetric swelling. When external stress is applied, the loop habit planes tilt to lower the elastic interaction energy\cite{wolfer2004motion}, resulting in a polarised eigenstrain\cite{Mason2020, boleininger2022volume}. This tilting is balanced by an increase in the dislocation line tension energy\cite{li2019diffusion}. Consequently, the degree of anisotropy increases with increasing external stress.

At higher doses, in the transient phase occurring between approximately 0.05 and {0.10\,dpa}, the loops grow and coalesce, eventually forming extended networks or even complete crystal planes\cite{Mason2020, boleininger2022volume}. The planes form preferentially normal to, or parallel to, the uniaxial tensile and compressive stress directions, respectively, as illustrated in Fig.~\ref{fig:5}a (plastic). Plane formation contributes significantly to the overall dimensional change. As dislocations structures transform into crystal planes, the counterbalancing effect of line tension vanishes, and their partially polarised relaxation volumes become fully polarised.

If the external stress is large, {$\lvert \sigma_\mathrm{zz}^\mathrm{ext} \rvert$\,$\gtrsim$\,1\,GPa}, the dislocation network can additionally undergo slip; in an individual atomistic simulation, this is manifested though a substantial change\cite{Derlet2020} in eigenstrain over a few millidpa. Among the three mechanisms, only slip can keep deforming the crystal indefinitely under continued irradiation, while loop tilting and plane formation eventually cease as the microstructure saturates with vacancies, and no further interstitials are created\cite{boleininger2023microstructure}.

The above relaxation mechanisms conserve volume, keeping the amount of swelling unchanged under application of external stress, see Fig.~\ref{fig:5}c. Examining the atomistic simulations, we find that mono-vacancies represent over {90\,\%} of the vacancy content, while interstitials coalesce into very large clusters. Therefore, the amount of swelling can be estimated using $\varepsilon^*_\mathrm{vol} = \mathrm{tr}(\tensor{\varepsilon}^*) \approx c_\mathrm{vac}\left(1 + \Omega_\mathrm{vac}\right)$, where $c_\mathrm{vac}$ represents the mono-vacancy number concentration, including those constituting vacancy clusters, and {$\Omega_\mathrm{vac}$\,$=$\,$-0.36$} is the relaxation volume of a tungsten mono-vacancy\cite{mason2017empirical} in atomic volumes, close to the \textit{ab initio} value\cite{MaPRM2019b} of {$\Omega_\mathrm{vac}$\,$=$\,$-0.345$}. The simulated microstructures saturate to a vacancy concentration of {$c_\mathrm{vac}$\,=\,(0.30\,$\pm$\,0.01)\,\%} with a volume swelling of {$\varepsilon^*_\mathrm{vol}$\,=\,(0.20$\,\pm$\,0.01)\,\%} (with uncertainty indicating standard deviation), which is in close agreement with the swelling estimate of {0.19\,\%} based on the vacancy concentration alone.

As was already hypothesised by Grossbeck and Mansur\cite{grossbeck1991low}, the mechanisms responsible for the anomalous, low-temperature irradiation creep are distinctly different from the conventional thermal irradiation creep mechanisms. Here, stress relaxation occurs entirely in the absence of thermal diffusion of defects. We observe that dislocation loops preferentially align themselves with respect to the external stress direction at low dose, see Fig.~\ref{fig:5}a at {0.04\,dpa}. This observation agrees with a recent study reaching a similar dose\cite{da2023evidence}, supporting the stress-induced preferential nucleation hypothesis. However, at higher doses, the dislocation microstructure evolves into a system-spanning network\cite{wang2023dynamic}, enabling the plastic deformation of the entire crystal\cite{boleininger2022volume}. This is correlated with the sudden increase in eigenstrain over this dose range, as seen in Fig.~\ref{fig:3}b, and a simultaneous decrease in the dislocation line density. This process is driven by the elastic interaction between dislocation loops, ultimately leading to their coalescence into an interconnected network. A microstructure with $c_\mathrm{vac} = \SI{0.3}{\percent}$ and preferentially tilted loops can at most attain $\varepsilon_\parallel^* = \SI{2.6e-3}{}$, which is already exceeded at $\sigma_\mathrm{zz}^\mathrm{ext} = \SI{0.5}{GPa}$, see Fig.~\ref{fig:3}b, illustrating that the crystal has undergone plastic deformation.

The peak eigenstrain value of {1\,\%} is quantitatively comparable to low-temperature irradiation creep measurements in steels\cite{grossbeck1991low}, where higher temperatures were found to yield \textit{lower} creep rates. 
Notably, we observed a significant amount of relaxation to occur within {0.3\,dpa} of exposure, which matches low-temperature \textit{in situ} irradiation creep experiments conducted in tungsten\cite{ponsoye1971irradiation}. We conclude that in a high-fluence reactor environment, stresses are expected to relax extremely quickly. Considering the operating conditions in a fusion power plant \cite{knaster2016materials}, it takes about a week of full-time operation to reach this dose in critically important components such as the first wall, divertor, or cooling pipes. While structural materials close to the plasma may be kept sufficiently hot to stimulate recombination of radiation defects, the actively cooled materials, in which stress concentrations critical to structural integrity are expected to develop, would relax through the favourable low-temperature creep mechanism explored in this study.

\bibliography{main.bbl}


\

\noindent\textbf{\sffamily  Methods}\\
{\footnotesize
\noindent\textbf{Stimulated stress relaxation experiment.}
The stimulated stress relaxation experiments were carried out using the \textbf{G}eneral-Purpose \textbf{Ir}r\textbf{a}diated \textbf{F}iber and \textbf{F}oil \textbf{E}xperiment (GIRAFFE) at the Max Planck Institute for Plasma Physics (IPP) in Garching. GIRAFFE is a specialised device designed and built for the study of synergistic effects in materials science, consisting of a tensile testing machine that can be integrated into one of the beamlines of a \SI{3}{MV} tandem accelerator.

The samples were fabricated out of potassium-doped, cold drawn tungsten wire with a diameter of \SI{16}{\um}, which was manufactured by \emph{\mbox{OSRAM} GmbH} in Germany. To prepare the samples, an approximately \SI{8}{cm} long piece of wire was affixed to a polypropylene (PP) frame using \emph{UHU PLUS ENDFEST 300} epoxy resin, which significantly improved the handling and  mounting of the sample. Defined electrical contacts were established at both ends using adhesive copper tape, leaving a central usable length of the sample measuring \SI{15}{mm}.

After the sample was mounted in the fixture of the tensile testing machine, it was loaded with about \SI{50}{mN}, and an x-y stage was translated between the sample fixture and the load cell in both axes until a minimum force was reached. This procedure ensured a uniaxial tensile load on the wire during the test. To measure the ion current, the lower sample holder was connected to a \emph{Keithley 6487} picoamperemeter. In addition, a tungsten beam-limiting aperture with a vertical opening of \SI{4}{mm} was installed, along with suppression electrodes maintained at a voltage of \SI{-500}{V} to minimise the effect of secondary electron emission from the sample and aperture. An additional \SI{102}{\um} thin 
tungsten wire was attached to the lower sample holder, which has no contact with the upper sample holder to avoid influencing the mechanical measurement. This wire performs two functions. During the experiment, it was positioned in the beam path behind the sample, thus increasing the measured ion current by a factor of $\frac{102}{16}$ due to its larger diameter, resulting in a significantly improved signal-to-noise ratio. The ion fluence ($\Phi$) was then calculated from the measured ion current ($I$), the ion charge state ($z$\,=\,6), the diameter of the thick wire ($d$\,=\,16\,{\textmu m}), the aperture height ($l_\mathrm{irr}$\,=\,4\,{mm}), and the elementary charge using equation
\begin{equation}
    \Phi(t) = \frac{1}{e Z d l_\mathrm{irr}}\int_{t_0}^{t} I(t') \, \textup{d}t'.
\end{equation}
The second function of the \SI{102}{\um} wire was to protect the sample during accelerator start-up and adjustment. For this purpose, the sample holders were rotated by \SI{180}{\degree} so that the sample lied in the beam shadow of the thick wire, preventing the sample from being irradiated before the start of the experiment.

Previous to the actual main experiment, a preliminary experiment was performed. In this test, the sample is mechanically tensioned to \SI{110}{\percent} of the initial value of the main experiment, and the force drop measured over a duration of about \SI{2}{h}. This was done to quantify the relaxation of the entire test setup. During the first \SI{30}{min}, an exponential drop of the force was observed, followed by a linear decay. This was likely due to settling effects in the adhesive bonding of the sample. For this reason, we delayed the start of the main experiment by \SI{30}{min} after the application of the mechanical load.

For the main experiment, the sample was again mechanically tensioned to initial values of either \SI{0.5}{GPa}, \SI{1.0}{GPa}, \SI{1.5}{GPa}, or \SI{2.0}{GPa}. After the aforementioned \SI{30}{min} delay, the irradiation with the high-energy heavy ion beam began, while the position of the sample ends is kept constant. Since almost all of the energy of the impacting ions was deposited in the sample, it heats up by less than \SI{100}{K}, resulting in thermal expansion and an immediate drop in the tensile force. To compensate for this, the irradiation was carried out in intervals, and the force drop was only measured during brief irradiation pauses after a short cooling phase. At the start of the experiment, the interval duration was \SI{5}{min} to provide higher time resolution, with the interval length gradually increased up to \SI{30}{min} during the experiment.

The \SI{20.3}{MeV} W$^{6+}$ ion beam used to generate the damage in the sample had a beam spot with a diameter of about \SI{1.5}{mm} and a Gaussian intensity distribution. To ensure uniform irradiation along the sample axis, the ion beam was scanned vertically at a frequency of \SI{1}{kHz} and limited to a height of \SI{4}{mm} using a tungsten aperture. Due to the significant diameter ratio between the ion beam and the sample wire, the intensity was assumed to be constant across the sample diameter. The net ion current passing through both wires was initially approximately \SI{0.15}{nA} to achieve high time resolution and was gradually increased up to \SI{0.9}{nA} during the course of the experiment to reach the desired final dose. This resulted in an ion flux in the range of \SIrange{2.5e11}{1.5e12}{cm^{-2} s^{-1}}.

The measured force values needed to be corrected for two sources of error. The first source was a long-term drift of the zero point in the load cell. A preparatory experiment showed that the zero point of the load cell shifted by less than \SI{1}{mN} during the typical 6 hour duration of the experiment. To quantify this effect, the sample was completely unloaded at the end of the main experiment at time $t_\mathrm{E}$, allowing the determination of the zero drift $F_\mathrm{E}$. Each measured value $F_i$ at time $t_i$ was then corrected using equation
\begin{equation}
    F_{i}^\mathrm{corr,1} = F_i - F_\mathrm{E} \frac{t_i}{t_\mathrm{E}}.
\end{equation}
The second correction pertained to the relaxation of the entire measurement setup, particularly the epoxy resin embedding of the sample ends. To address this, the stress-strain curve obtained from the preliminary experiment was divided into intervals of 1000 data points, and the slope of each interval was determined. The individual slopes were then divided by the average force in the interval, and the overall slope average ($\alpha$) was calculated. The measured values were then corrected using equation
\begin{equation}
    F_{i}^\mathrm{corr,2} = \sum_{j=1}^{i}(t_j - t_{j-1}) \cdot \alpha \cdot \frac{F_j + F_{j-1}}{2}.
\end{equation}

\

\noindent\textbf{Molecular dynamics.} Simulations were run using the LAMMPS\cite{plimpton1995fast} software, employing an embedded atom model potential for tungsten\cite{mason2017empirical}. This potential was chosen because it accurately predicts the defect content of highly irradiated material, in agreement with experiments\cite{mason2021parameter, boleininger2023microstructure}. All simulations were initialised as pristine single-crystal tungsten consisting of 100$\times$100$\times$250 body-centered crystal unit cells ({32$\times$32$\times$80\,nm$^3$}), with periodic boundary conditions applied to all three directions in order to describe a materials volume element representative of a grain in a wire. Uniaxial stress was applied along the $z$-direction by minimising the potential energy of the system with respect to atom positions and box dimensions using the method of conjugate gradients, under the constraint that all but the $\sigma_\mathrm{zz}$ component of the internal stress tensor reach near zero. To introduce irradiation damage, rather than simulating the entire track of incident high-energy tungsten ions that extend several microns into the material, we leveraged the known property of collision sequences in which damage-generating scattering events are relatively rare and can be considered as independent and uncorrelated. We used the \textsc{srim} software\cite{ziegler2004srim}, which is based on the binary collision approximation method, to generate a set of primary recoil energies representative of those produced by {20.3\,MeV} W$^{6+}$ ions impacting solid bulk tungsten. The effect of irradiation was then introduced by the following iterative procedure: First, recoil energies are randomly drawn from the representative set of recoil energies until a target dose increment of $\sum_{i} N_\mathrm{d}(T_\mathrm{d}^{(i)})/N$\,$>$\,{0.0002\,dpa} is reached, where $N_\mathrm{d}(T_\mathrm{d})$ is an estimate for the number of atoms displaced by a recoil with damage energy $T_\mathrm{d}$ in the NRT model\cite{norgett1975proposed}; a convention commonly used in the nuclear materials community. The damage energy $T_\mathrm{d}^{(i)}$ is computed from the recoil energy $E_R^{(i)}$ using Lindhard's formula\cite{norgett1975proposed} to account for losses due to electronic stopping. For each drawn recoil energy, a random atom in the system is selected and assigned the corresponding kinetic energy with randomly-oriented velocity. We avoided selecting atoms too close to one another to prevent any spurious effects caused by overlapping regions where damage production occurs. Next, the simulation is propagated in the NVE ensemble for a duration of {10\,ps}, which was sufficient for the cascade dynamics to conclude. We including damping terms describing energy loss due to electronic excitations; for more details on this, we refer to Ref.\cite{boleininger2023microstructure}. After the propagation, the atom velocities are set to zero, and the system is relaxed to a local energy minimum while maintaining the uniaxial stress constraint. This procedure is repeated, with every iteration incrementing the dose by approximately {0.0002\,dpa}, until a target dose of {0.5\,dpa} is reached. After each relaxation step, the simulation cell dimensions were recorded in a file, from which we computed the eigenstrain as a function of dose for the specified target stress using the formula $\tensor{\varepsilon}^*(\phi;\sigma_\mathrm{zz}) = \tensor{A}_i \cdot \tensor{A}_0^{-1} - \mathds{I}$, where $\tensor{A}_i$ is a matrix with columns represented by the simulation cell vectors at dose $\phi_i$.

\

\noindent\textbf{Virtual wire model.} The full boundary value problem is given by
\begin{equation}\label{eq:totalelastostatic}
\begin{aligned}
\vec{\nabla}\cdot \tensor{\sigma}(\vec{r},t)              
    &= \vec{\nabla}\cdot \tensor{\sigma}^*(\vec{r},t)
        &\quad\quad  &\vec{r} \in \Omega\\ 
\tensor{\sigma}(\vec{r},t) \cdot \vec{n}(\vec{r}) 
    &= \tensor{\sigma}^*(\vec{r},t)
        \cdot \vec{n}(\vec{r})
        &\quad\quad  &\vec{r} \in \partial\Omega_\mathrm{t}\\ 
\vec{u}(\vec{r},t) &= \vec{u}^\mathrm{C}(\vec{r}) 
        &\quad\quad  &\vec{r} \in \partial\Omega_\mathrm{u}.
\end{aligned}
\end{equation}
We solve for the displacement field $\vec{u}(\vec{r})$ within the region $\Omega \subset \mathbb{R}^3$ representing the elastic body, subject to traction and displacement conditions applied to the surfaces $\partial\Omega_t$ and $\partial\Omega_u$, respectively, with $\partial\Omega_t \cup \partial\Omega_u = \partial\Omega$. Vector field $\vec{n}(\vec{r})$ represents the outward-facing normal vector of surface $\partial\Omega$. 

The boundary value problem was solved using the finite element method (FEM) with the FEniCS\cite{AlnaesEtal2015} solver. In practice, we exploit that the wire diameter ($d$\,=\,16\,{\textmu m}) is significantly smaller than both the initial wire length ($l$\,=\,15\,{mm}) and the length of the irradiated wire section  ($l_\mathrm{irr}$\,=\,4\,{mm}), and therefore only solve for the fields within a two-dimensional x-y slice of the irradiated wire section, considering it as representative of the entire irradiated region. In the simulations, we described the uniaxial tensile loads applied in experiment as an external homogeneous stress $\tensor{\sigma}^\mathrm{ext}$, in which only the $\sigma^\mathrm{ext}_\mathrm{zz}$ component was non-zero. The wire boundaries were subject to traction free conditions. Before simulating the stimulated stress relaxation, we required a spatial map of the dose-rate $\dot{\phi}(x,y)$ within the wire slice, originating by the ion beam irradiation, as the wire is only irradiated from one side and the beam does not penetrate beyond a surface layer of about {2\,\textmu m} thickness. We used \textsc{srim}\cite{ziegler2004srim} to generate the dose-rate profile $\dot{\phi}(x,y)$ shown in Fig.~\ref{fig:1}, with more comprehensive details given in the supplemental material. Beginning with step $n=0$, the initial dose and eigenstrain were set to zero: $\phi_n(x,y) = 0$ and $\tensor{\varepsilon}^*_n(x,y) = 0$. Additionally, the initial external stress was set at $\tensor{\sigma}^\mathrm{ext}_0(x,y) = \tensor{\sigma}^\mathrm{ext}$. The stress relaxation was propagated using the following explicit method: First, the FEM problem is solved for the total stress $\tensor{\sigma}_n(x,y)$, from which we obtain the elastic stress in the irradiated region by subtracting the eigenstrain contribution\cite{Reali2022},
$\tensor{\sigma}^\mathrm{el}_n(x,y) = \tensor{\sigma}_n(x,y) - \tensor{C} : \tensor{\varepsilon}^*_n(x,y)$.
Next, the dose profile is updated to the next time step, 
$\phi_{n+1}(x,y) = \phi_{n}(x,y) + \dot{\phi}(x,y)\Delta t$,
and the eigenstrain is incremented using the previous elastic stress state, under assumption that dimensional changes are accrued irreversibly:
\begin{equation*}
\tensor{\varepsilon}^*_{n+1}(x,y) = \tensor{\varepsilon}^*_n(x,y) 
    + \tensor{\varepsilon}_\mathrm{MD}^*[\phi_{n+1}, \tensor{\sigma}_n^\mathrm{el}](x,y)
    - \tensor{\varepsilon}_\mathrm{MD}^*[\phi_n, \tensor{\sigma}_n^\mathrm{el}](x,y).
\end{equation*}
The new eigenstrain is used to update body forces and surface tractions for the next iteration of the FEM solver. Relaxation of the uniaxial stress arises from the length change of the irradiated section, which follows entirely from the eigenstrain\cite{Reali2022} through
$\langle \sigma^\mathrm{el}_{\mathrm{zz}}\rangle  = \sigma^\mathrm{ext}_{\mathrm{zz}} - E \langle \varepsilon_{\mathrm{zz},n}^*\rangle l_\mathrm{irr}/l$, 
where $E$ is the Young's modulus of tungsten and $\langle \varepsilon^*_{\mathrm{zz},n} \rangle$ is the component of the eigenstrain along the wire length at step $n$, averaged over the wire cross-section. This procedure is repeated until the simulation advances to the desired exposure time. After every iteration, the profiles of dose, eigenstrain, and elastic stress are exported.

\

\noindent\textbf{Uncertainty quantification.} The stochastic variation of the atomistic eigenstrain data-sets gives rise to uncertainty on the surrogate model parameters  $\hat{\vec{\theta}}$, which propagates up the scale as uncertainties on the predictions from the virtual wire model. Using the Metropolis-Hastings (MH) algorithm, we obtain $10^5$ random samples of the parameter vector $\vec{\theta}$. The Markov chain is started with an initial value of $\vec{\theta}_0 = \hat{\vec{\theta}}$, and moves are proposed according to the protocol $\vec{\theta}_{n+1} = \vec{\theta}_n + x\cdot \vec{\mathcal{N}}(0, \vec{v})$, where $\vec{v}$ is a vector with components $v_i = \hat{s}_i^2$ and $x$ is a step-size between 0.02 and 0.03, chosen to maintain an acceptance ratio between \SI{10}{\percent} and \SI{20}{\percent}. A move is accepted if $r \leq \mathcal{L}(\vec{\theta}_{n+1} | \vec{\varepsilon}^*)/\mathcal{L}(\vec{\theta}_{n} | \vec{\varepsilon}^*)$, where $r \in [0,1]$ is a uniform random number. The initial $10^4$ samples are discarded to allow for burn-in. This process is repeated for every eigenstrain data-set for each common external stress, resulting in $10^5$ realisations of surrogate model parameter vectors $\vec{\theta}$ per stress for both parallel and perpendicular eigenstrains. Finally, every virtual wire simulation is independently executed $\SI{5e3}{}$ times, each time using a surrogate model $\tensor{\varepsilon}^*_\mathrm{MD}(\phi, \sigma_\mathrm{zz}^\mathrm{ext})$ with parameter vectors randomly drawn from their respective Markov chains. From this, a distribution of stress relaxation curves is obtained, allowing construction of the confidence intervals shown in Figs.~\ref{fig:1} and \ref{fig:4}.
}

\

\noindent\textbf{\sffamily  Data availability}\\
{\footnotesize
The datasets generated during the current study will be made public on acceptance of this paper.
}

\

\noindent\textbf{\sffamily  Code availability}\\
{\footnotesize
The open-source computer code LAMMPS used for MD simulations is available at \url{https://lammps.sandia.gov}. The open-source Python package FEniCS used for multiscale simulations is available at \url{https://fenicsproject.org/}.
}


\

\noindent\textbf{\sffamily  Acknowledgements}\\
{\footnotesize
We are grateful to Robert Lürbke for taking the photograph shown in Fig.~\ref{fig:1}a. This work has been carried out within the framework of the EUROfusion Consortium, funded by the European Union via the Euratom Research and Training Programme (Grant Agreement No. 101052200 - EUROfusion), and by the RCUK Energy Programme, Grant No. EP/W006839/1. To obtain further information on the data and models underlying the paper please contact PublicationsManager@ukaea.uk. The views and opinions expressed herein do not necessarily reflect those of the European Commission. The authors acknowledge the use of the Cambridge Service for Data Driven Discovery (CSD3) and associated support services provided by the University of Cambridge Research Computing Services \cite{csd3} that assisted the completion of this study.
}

\

\noindent\textbf{\sffamily Author contributions}\\
{\footnotesize
A.F. and M.B. contributed equally to the formulation of the scientific problem and conceptualisation of experimental and computational studies. A.F. designed and conducted the experiments and analysed the stress relaxation data. M.B. developed the multiscale simulation framework and conducted the simulations. S.L.D. and J.R. contributed jointly to design of the research project. All authors contributed to the data discussion and manuscript preparation.}

\

\noindent\textbf{\sffamily Competing interests}\\
{\footnotesize
The authors declare no competing interests.
}

\end{document}